\documentclass[journal, draftcls,onecolumn]{IEEEtran}
\usepackage{enumerate}
\usepackage{epstopdf}
\usepackage{cite}
\usepackage{amscd}
\usepackage{dsfont}
\usepackage{latexsym}
\usepackage{array}

\usepackage{tikz}\usetikzlibrary{calc}
\usepackage{tabularx}
\usepackage{epsfig}
\usepackage{graphicx,subfigure}
\usepackage{epstopdf}
\usepackage{graphicx}
\usepackage{amsmath,mathrsfs}
\usepackage{amsthm}
\usepackage{amssymb}  
\usepackage{amsbsy}
\usepackage{color}
\usepackage{stfloats}
\usepackage{algorithm}
\usepackage{algpseudocode}
\usepackage{bm}
\usepackage{pifont}
\usepackage{accents}
\usepackage{multirow}

\usepackage{etoolbox}
\makeatletter
\patchcmd{\@makecaption}
  {\scshape}
  {}
  {}
  {}
\makeatother

\newtheoremstyle{note}
{3pt}
{1pt}
{}
{\parindent}
{\itshape}
{:}
{.5em}
{\thmname{#1}\thmnumber{ #2}\thmnote{\thmnote{ (#3)}}}
\theoremstyle{note}

\newtheorem{definition}{Definition}

\newtheorem{ther}{Theorem}

\theoremstyle{definition}
\newtheorem{rem}{Remark}
\newtheoremstyle{dotless}{}{}{\itshape}{}{\bfseries}{}{ }{}
\theoremstyle{dotless}

\newcommand {\aplt} {\ {\raise-.5ex\hbox{$\buildrel<\over{\mbox{\scriptsize $\sim$}}$}}\ }

\begin{document}
%
\title{Polar Codes and Polar Lattices for Independent Fading Channels}
\author{Ling~Liu
	    and Cong Ling~\IEEEmembership{Member,~IEEE}
    \thanks{This work was supported in part by Huawei's Shield Lab through the HIRP Flagship Program and in part by the China Scholarship Council.}
	\thanks{Ling Liu and Cong Ling are with the Department of Electrical and Electronic Engineering,
	 Imperial College London, London, UK (e-mails: l.liu12@imperial.ac.uk, cling@ieee.org).}
}

\maketitle
\begin{abstract}
In this paper, we design polar codes and polar lattices for i.i.d. fading channels when the channel state information is only available to the receiver. For the binary input case, we propose a new design of polar codes through single-stage polarization to achieve the ergodic capacity. For the non-binary input case, polar codes are further extended to polar lattices to achieve the egodic Poltyrev capacity, i.e., the capacity without power limit. When the power constraint is taken into consideration, we show that polar lattices with lattice Gaussian shaping achieve the egodic capacity of fading channels. The coding and shaping are both explicit, and the overall complexity of encoding and decoding is $O(N \log^2 N)$.
\end{abstract}

\IEEEpeerreviewmaketitle

\section{Introduction}
Real-world wireless channels are generally modeled as time-varying fading channels due to multiple signal paths and user mobility. Compared with time-invariant channel models, the wireless fading channel models allow the channel gain to change randomly over time. In practice, we usually consider slow and fast fading channels. In slow fading channels, the channel gain varies at a larger time scale than the symbol duration. In fast fading channels, the code block length typically spans a large number of coherence time intervals and the channel is ergodic with a well-defined Shannon capacity. In this paper we study the fast fading channel with independent channel gains. This may be realized by perfect interleaving/de-interleaving of symbols, which offers much convenience for design. We further assume that channel state information (CSI) is available to the receiver through training sequences, and the transmitter only has the channel distribution information (CDI).

Polar codes, introduced by Ar{\i}kan \cite{arikan2009channel}, are capacity achieving for binary-input memoryless symmetric channels (BMSCs). Efficient construction methods of polar codes for classical BMSCs such as binary erasure channels (BECs), binary symmetric channels (BSCs), and binary-input additive white Gaussian noise (BAWGN) channels were proposed in \cite{Ido,mori2009performance,PolarConstru}. Besides channel coding, polar codes were then extended to source coding and their asymptotic performance was proved to be optimal \cite{polarsource,KoradaSource}. As a combination of the application of polar codes for channel coding and lossless source coding, polar codes were further studied for binary-input memoryless asymmetric channels (BMACs) in \cite{aspolarcodes,howtoachieveAsym,PolarAsymOrig}. The versatility of polar codes makes them attractive and promising for coding over many other channels, such as wiretap channels \cite{polarsecrecy}, broadcast channels \cite{polarbroadcast}, multiple access channels (MACs) \cite{AbbeMac}, compound channels \cite{PolarUniHassani,PolarUniSasoglu} and even quantum channels \cite{PolarQuantum}.

For fading channels, there has been considerable progress. Quasi-static fading channel with two states was studied in \cite{PolarFadingBoutros}. Construction of polar codes for block Rayleigh fading channels when CSI or CDI is available to both transmitter and receiver was considered in \cite{PolarFadingAngel}. In this work, we consider the case in which CSI is available to the receiver, and the transmitter only knows CDI. This is the case when a communication system is operated in the open-loop mode. We show that the same channel capacity can be achieved as in the case where CSI is available to both. The previous work \cite{PolarFadingHongBo} of polar codes for fading channels does not require CSI for the transmitter either. The authors proposed a novel hierarchic scheme to construct polar codes through two phases of polarization. The channel state is assumed to
be constant over each coherence interval and the channel is modeled as a mixture of BSCs. The first phase of polarization is to get each BSC polarized into a set of extremal subchannels (ignoring the unpolarized part), which is treated as a set of realizations of BECs. Then the second phase of polarization is to get the synthesized BECs polarized. This scheme achieves the ergodic capacity of binary input fading channels with finite states when the two phases are both sufficiently polarized. As a result, much longer block length than standard polar codes is needed to achieve channel capacity. In this paper, we propose a new scheme with one-phase polarization to achieve the ergodic capacity by treating the channel gain as part of channel outputs.


As the counterpart of linear codes in the Euclidean space, lattice codes provide more freedom over signal constellation for communication systems. The existence of lattice codes achieving the point-to-point additive white Gaussian noise (AWGN) channel capacity was established in \cite{zamir,LingBel13}. Besides point-to-point communications, lattice codes are also useful in a wide range of applications in multiterminal communications, such as information-theoretical security \cite{cong2}, compute-and-forward \cite{nazer}, distributed source coding \cite{zamir1}, and $K$-user interference channel \cite{OrdentlichInterfe} (see \cite{BK:Zamir} for an overview). The two important ingredients of AWGN capacity-achieving lattice coding are AWGN-good lattices \cite{zamir} and shaping. Following the work on multilevel codes \cite{forney6}, polar lattices were constructed from polar codes according to ``Construction D" \cite{yellowbook} and proved to be AWGN-good \cite{yan2}. With lattice Gaussian shaping \cite{LingBel13}, polar lattices were then shown to be capable of achieving the AWGN capacity \cite{polarlatticeJ}. More recently, random lattice codes were investigated in ergodic fading channels \cite{HindyISIT}. However, the explicit construction of lattice codes for ergodic fading channels is an open problem. In this work, we will resolve this problem using polar lattices for i.i.d. fading channels.

For fading channels, algebraic tools \cite{BK:fading} play an important role in explicit coding design. It was shown in \cite{AlgebraicFading} that lattice codes constructed from algebraic number field can achieve full diversity over fading channels, which results in better error performance. A more recent work showed that number field lattices are able to achieve the Gaussian and the Rayleigh channel capacity within a constant gap \cite{LuzziNumLattice}. This scheme is universal and extended to the multi-input and multi-output (MIMO) context \cite{LuzziNumMIMO}.

The paper is organized as follows:
Section II presents the background of polar codes and polar lattices. The construction of polar codes for binary-input i.i.d. fading channels is investigated in Section III, along with some simulation results.
In Section IV, we firstly design polar lattices for fading channels without power constraint and prove that the ergodic Poltyrev capacity can be achieved; lattice Gaussian shaping is then implemented to obtain the optimum shaping gain. Finally, the paper is concluded in Section V.

All random variables (RVs) are denoted by capital letters. Let $P_X$ denote the probability distribution of a RV $X$ taking values $x$ in a set $\mathcal{X}$. For multilevel coding, we denote by $X_\ell$ a RV $X$ at level $\ell$. The $i$-th realization of $X_\ell$ is denoted by $x_\ell^i$. We also use the notation $x_\ell^{i:j}$ as a shorthand for a vector $(x_\ell^i,..., x_\ell^j)$, which is a realization of RVs $X_\ell^{i:j}=(X_\ell^i,..., X_\ell^j)$. Similarly, $x_{\ell:\jmath}^i$ denotes the realization of the $i$-th RV from level $\ell$ to level $\jmath$, i.e., of $X_{\ell:\jmath}^i=(X_\ell^i,..., X_\jmath^i)$. For a set $\mathcal{I}$, $\mathcal{I}^c$ denotes its complement, and $|\mathcal{I}|$ represents its cardinality. For an integer $N$, $[N]$ will be used to denote the set of all integers from $1$ to $N$. Following the notation of \cite{arikan2009channel}, we denote $N$ independent uses of channel $W$ by $W^N$. By channel combining and splitting, we get the combined channel $W_N$ and the $i$-th subchannel $W_N^{(i)}$. The binary logarithm and natural logarithm are accordingly denoted by $\log$ and $\ln$, and information is measured in bits.

\section{Preliminaries of Polar Codes and Polar Lattices}

\subsection{Polar Codes}
Let $\tilde{W}$ be a BMSC with input alphabet $X \in \mathcal{X}=\{0,1\}$ and output alphabet $Y \in \mathcal{Y}\subseteq\mathbb{R}$. Given the capacity $C(\tilde{W})$ of $\tilde{W}$ and a rate $R<C(\tilde{W})$, the information bits of a polar code with block length $N=2^m$ are indexed by a set of $\lfloor RN \rfloor$ rows of the generator matrix $G_N=\left[\begin{smallmatrix}1&0\\1&1\end{smallmatrix}\right]^{\otimes m}$, where $\otimes$ denotes the Kronecker product. The matrix $G_N$ combines $N$ identical copies of $\tilde{W}$ to $\tilde{W}_N$. Then this combination can be successively split into $N$ binary memoryless symmetric subchannels, denoted by $\tilde{W}_{N}^{(i)}$ with $1 \leq i \leq N$. By channel polarization, the fraction of good (roughly error-free) subchannels is about $C(\tilde{W})$ as $m\rightarrow \infty$. Therefore, to achieve the capacity, information bits should be sent over those good subchannels and the rest are fed with frozen bits which are known before transmission. The indices of good subchannels can be identified according to their associated Bhattacharyya Parameters.

\begin{definition}[Bhattacharyya Parameter for Symmetric Channel \cite{arikan2009channel}]\label{deft:symZ}
Given a BMSC $\tilde{W}$ with transition probability $P_{Y|X}$, the Bhattacharyya parameter $\tilde{Z}\in[0,1]$ is defined as
\begin{eqnarray}
\tilde{Z}(\tilde{W})&\triangleq\sum\limits_{y} \sqrt{P_{Y|X}(y|0)P_{Y|X}(y|1)}.
\end{eqnarray}
\end{definition}

Based on the Bhattacharyya parameter, the information set $\tilde{\mathcal{I}}$ is defined as $\{i:\tilde{Z}(\tilde{W}_{N}^{(i)})\leq 2^{-N^{\beta}}\}$ for some $0<\beta <\frac{1}{2}$, and the frozen set $\tilde{\mathcal{F}}$ is the complement of $\tilde{\mathcal{I}}$. Let $P_B$ denote the block error probability of a polar code under successive cancellation (SC) decoding. It can be upper-bounded as $P_{B}\leq\Sigma_{i\in \tilde{\mathcal{I}}}\tilde{Z}(\tilde{W}_{N}^{(i)})$. An efficient algorithm to evaluate the Bhattacharyya parameter of subchannels for general BMSCs was presented in \cite{Ido,PolarConstru}.

The following definition of channel degradation will be frequently used.

\begin{definition}[Channel degradation]\label{deft:Degrade}
Let $W_1:X \to Y_1$ and $W_2: X \to Y_2$ be two channels. $W_1$ is stochastically degraded with respect to $W_2$ if there exists an intermediate channel $W:Y_2\rightarrow Y_1$ such that
\begin{eqnarray}
W_1(y_1|x)=\sum_{y_2\in\mathcal{Y}_2}W_2(y_2|x)W(y_1|y_2).
\end{eqnarray}
\end{definition}

\begin{rem}\label{rk:polardegrade}
Let $\tilde{W}$ and $\tilde{V}$ be two BMSCs. If $\tilde{V}$ is degraded with respect to $\tilde{W}$, after channel polarization, $\tilde{Z}(\tilde{W}_N^{(i)}) \leq \tilde{Z}(\tilde{V}_N^{(i)})$, and the polar code $C_V$ constructed according to the Bhattacharyya parameter rule for $\tilde{V}$ is a subcode of the polar code $C_W$ for $\tilde{W}$, i.e., $C_V \subseteq C_W$ \cite{polarchannelandsource}.
\end{rem}

\subsection{Lattice Codes}
An $n$-dimensional lattice is a discrete subgroup of $\mathbb{R}^{n}$ which can be described by
\begin{eqnarray}
\Lambda=\{ \lambda=\mathbf{B}z:z\in\mathbb{Z}^{n}\},
\end{eqnarray}
where the columns of the generator matrix $\mathbf{B}=[\mathrm{b}_{1}, \cdots, \mathrm{b}_{n}]$ are assumed to be linearly independent.

For a vector ${x}\in\mathbb{R}^{n}$, the nearest-neighbor quantizer associated with $\Lambda$ is $Q_{\Lambda}({x})=\text{arg}\min\limits_{ \lambda\in\Lambda}\|\lambda-{x}\|$. We define the modulo lattice operation by ${x} \text{ mod }\Lambda\triangleq {x}-Q_{\Lambda}({x})$. The Voronoi region of $\Lambda$, defined by $\mathcal{V}(\Lambda)=\{{x}:Q_{\Lambda}({x})=0\}$, specifies the nearest-neighbor decoding region. The Voronoi region is one example of the fundamental region of a lattice. A measurable set $\mathcal{R}(\Lambda)\subset\mathbb{R}^{n}$ is a fundamental region of the lattice $\Lambda$ if $\cup_{\lambda\in\Lambda}(\mathcal{R}(\Lambda)+\lambda)=\mathbb{R}^{n}$ and if $(\mathcal{R}(\Lambda)+\lambda)\cap(\mathcal{R}(\Lambda)+\lambda')$ has measure 0 for any $\lambda\neq\lambda'$ in $\Lambda$. The volume of a fundamental region is equal to that of the Voronoi region $\mathcal{V}(\Lambda)$, which is given by $V(\Lambda)=|\text{det}({B})|$.

For an $n$-dimensional lattice $\Lambda$, define the volume-to-noise ratio (VNR) by
\begin{eqnarray}
\gamma_{\Lambda}(\sigma)\triangleq\frac{V(\Lambda)^\frac{2}{n}}{\sigma^2}.
\end{eqnarray}

For $\sigma>0$ and $c\in\mathbb{R}^{n}$, we define the Gaussian distribution of variance $\sigma^{2}$ centered at $c$ as
\begin{eqnarray}
f_{\sigma,c}(x)=\frac{1}{(\sqrt{2\pi}\sigma)^{n}}e^{-\frac{\| x-c\|^{2}}{2\sigma^{2}}}, \:\:x\in\mathbb{R}^{n}.
\end{eqnarray}

Let $f_{\sigma,0}(x)=f_{\sigma}(x)$ for short. For an AWGN channel with noise variance $\sigma^2$ per dimension, the probability of error $P_e(\Lambda, \sigma^2)$ of a minimum-distance decoder for $\Lambda$ is
\begin{eqnarray}
P_e(\Lambda, \sigma^2)=1-\int_{\mathcal{V}(\Lambda)} f_{\sigma}(x) dx.
\end{eqnarray}

\begin{definition}[AWGN-good lattices]\label{deft:awgngood}
A sequence of lattices $\Lambda^{(n)}$ of increasing dimension $n$ is AWGN-good if, for any fixed $P_{e}(\Lambda^{(n)},\sigma^2)\in(0,1)$,
\begin{eqnarray}
\lim_{n\rightarrow\infty}\gamma_{\Lambda^{(n)}}(\sigma)=2\pi e.
\end{eqnarray}
\end{definition}

The $\Lambda$-periodic function is defined as
\begin{eqnarray}
f_{\sigma,\Lambda}(x)=\sum\limits_{\lambda\in\Lambda}f_{\sigma,\lambda}(x)=\frac{1}{(\sqrt{2\pi}\sigma)^{n}}\sum\limits_{\lambda\in\Lambda}e^{-\frac{\| x-\lambda\|^{2}}{2\sigma^{2}}}.
\end{eqnarray}
We note that $f_{\sigma,\Lambda}(x)$ is a probability density function (PDF) if $x$ is restricted to the fundamental region $\mathcal{R}(\Lambda)$. This distribution is actually the PDF of the $\Lambda$-aliased Gaussian noise, i.e., the Gaussian noise after the mod-$\Lambda$ operation \cite{forney6}.

The flatness factor of a lattice $\Lambda$ is defined as \cite{cong2}
\begin{eqnarray}
\epsilon_{\Lambda}(\sigma)\triangleq\max\limits_{x\in\mathcal{R}(\Lambda)}| V(\Lambda)f_{\sigma,\Lambda}(x)-1|.
\end{eqnarray}

\begin{rem}\label{rk:flatness}
$\epsilon_{\Lambda}(\sigma_1) < \epsilon_{\Lambda}(\sigma_2)$, if $\sigma_1 > \sigma_2$ \cite{cong2}.
\end{rem}

We define the discrete Gaussian distribution over $\Lambda$ centered at $c$ as the discrete distribution taking values in $\lambda \in \Lambda$:
\begin{eqnarray}\label{eq:latticeGaussian}
D_{\Lambda,\sigma,c}(\lambda)=\frac{f_{\sigma,c}(\lambda)}{f_{\sigma,c}(\Lambda)}, \; \forall \lambda \in \Lambda,
\end{eqnarray}
where $f_{\sigma,\mathrm{c}}(\Lambda)=\sum_{\lambda \in \Lambda} f_{\sigma,\mathrm{c}}(\lambda)$. For convenience, we write $D_{\Lambda,\sigma}=D_{\Lambda,\sigma,\mathrm{0}}$. It has been proved to achieve the optimum shaping gain when the flatness factor is negligible \cite{LingBel13}.

A sublattice $\Lambda' \subset \Lambda$ induces a partition (denoted by $\Lambda/\Lambda'$) of $\Lambda$ into equivalence groups modulo $\Lambda'$. The order of the partition is denoted by $|\Lambda/\Lambda'|$, which is equal to the number of the cosets. If $|\Lambda/\Lambda'|=2$, we call this a binary partition. Let $\Lambda(\Lambda_0)/\Lambda_{1}/\cdots/\Lambda_{r-1}/\Lambda' (\Lambda_{r})$ for $r \geq 1$ be an $n$-dimensional lattice partition chain. If only one level is applied ($r=1$), the construction is known as ``Construction A". If multiple levels are used, the construction is known as ``Construction D" \cite[p.232]{yellowbook}. For each
partition $\Lambda_{\ell-1}/\Lambda_{\ell}$ ($1\leq \ell \leq r$) a code $C_{\ell}$ over $\Lambda_{\ell-1}/\Lambda_{\ell}$
selects a sequence of coset representatives $a_{\ell}$ in a set $A_{\ell}$ of representatives for the cosets of $\Lambda_{\ell}$. This construction requires a set of nested linear binary codes $C_{\ell}$ with block length $N$ and dimension of information bits $k_{\ell}$, which are represented as
$[N,k_{\ell}]$ for $1\leq\ell\leq r$ and $C_{1}\subseteq C_{2}\cdot\cdot\cdot\subseteq C_{r}$. Let $\psi$ be the natural embedding of $\mathbb{F}_{2}^{N}$ into $\mathbb{Z}^{N}$, where $\mathbb{F}_{2}$ is the binary field. Consider $\mathrm{g}_{1}, \mathrm{g}_{2},\cdots, \mathrm{g}_{N}$ be a basis of $\mathbb{F}_{2}^{N}$ such that $\mathrm{g}_{1},\cdots \mathrm{g}_{k_{\ell}}$ span $C_{\ell}$. When $n=1$, the binary lattice $L$ consists of all vectors of the form
\begin{eqnarray}
\sum_{\ell=1}^{r}2^{\ell-1}\sum_{j=1}^{k_{\ell}}\alpha_{j}^{(\ell)}\psi(\mathrm{g}_{j})+2^{r}z,
\label{constructionD}
\end{eqnarray}
where $\alpha_{j}^{(\ell)}\in\{0,1\}$ and $z\in\mathbb{Z}^{N}$. When $\{C_1,...,C_r\}$ is a series of nested polar codes, we obtain a polar lattice \cite{yan2}.

\section{Polar Codes for Binary-input fading channels}\label{sec:bfading}
Consider the binary-input i.i.d. fading channel
\begin{eqnarray}
Y=HX+Z,
\end{eqnarray}
where $X\in \{-1,+1\}$ is the binary input signal after BPSK modulation, $Y$ is the channel output, $Z$ is a zero mean independent Gaussian noise with variance $\sigma^2$, and $H$ is the channel gain. In this work, for convenience, we assume that $H$ follows the Rayleigh distribution with PDF
\begin{eqnarray}
P_H(h)=\frac{h}{\sigma_h^2} e^{-\frac{h^2}{2\sigma_h^2}},
\end{eqnarray}
where the scale parameter $\sigma_h=\sqrt{\frac{2}{\pi}} \cdot E[H]$. Denote by $SNR=\frac{\sigma_h^2}{\sigma^2}$ the signal noise ratio. Note that our work can be easily generalized to other regular fading distributions \cite{ShlomiFading}.

\begin{figure}[ht]
    \centering
    \includegraphics[width=6cm]{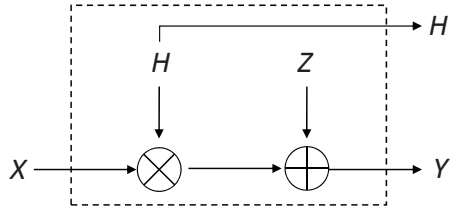}
    \caption{Binary-input fading channel with CSI available at the receiver.}
    \label{fig:bfading}
\end{figure}

Since we assume that $H$ is available to the receiver, the fading channel can be modeled as a channel with input $X$ and outputs $(Y,H)$, as shown in Fig. \ref{fig:bfading}.

We firstly show that the channel $\tilde{W}: X \to (Y,H)$ is symmetric. To see this, we check the channel transition PDF of $\tilde{W}$, which is given by
\begin{eqnarray}\label{eq:FadDens}
\begin{aligned}
P_{Y,H|X}(y,h|x)&=P_H(h)P_{Y|X,H}(y|x,h) \\
&=P_H(h)P_{Z}(z=y-xh) \\
&=P_H(h) \frac{1}{\sqrt{2 \pi \sigma^2}} e^{-\frac{(y-xh)^2}{2\sigma^2}}.
\end{aligned}
\end{eqnarray}
We define a permutation $\phi$ over the outputs $(y,h)$ such that $\phi(y,h)=(-y,h)$. Check that $P_{Y,H|X}(y,h|+1)=P_{Y,H|X}(\phi(y,h)|-1)$ and hence $\tilde{W}$ is symmetric. It is well-known that uniform input distribution achieves the capacity of symmetric channels. Therefore, letting $X$ be uniform, the capacity of $\tilde{W}$ is given by
\begin{eqnarray}
\begin{aligned}
C(\tilde{W})&=I(X;Y,H) \\
&=I(X;Y|H)\\
&=\sum_{x} \int_{0}^{\infty} \frac {h}{\sigma_h^2} e^{-\frac{h^2}{2\sigma_h^2}}dh \int_{-\infty}^{\infty} \frac{1}{2}\frac{1}{\sqrt{2 \pi \sigma^2}} e^{-\frac{(y-xh)^2}{2\sigma^2}} \log\Bigg(\frac{e^{-\frac{(y-xh)^2}{2\sigma^2}}}{\frac{1}{2}e^{-\frac{(y-h)^2}{2\sigma^2}}+\frac{1}{2}e^{-\frac{(y+h)^2}{2\sigma^2}}}\Bigg)dy\\
&=1-\frac{1}{\sqrt{2 \pi }\sigma \sigma_h^2} \int_{0}^{\infty} h e^{-\frac{h^2}{2\sigma_h^2}}dh \int_{-\infty}^{\infty} \Big(1-\log\Big(1+e^{-\frac{2yh}{\sigma^2}}\Big)\Big) dy,
\end{aligned}
\end{eqnarray}
which is the same as the capacity when the CSI is available to both transmitter and receiver \cite{PolarFadingAngel}.

To achieve $C(\tilde{W})$, we combine $N$ independent copies of $\tilde{W}$ to $\tilde{W}_N$ and split it to obtain subchannel $\tilde{W}_N^{(i)}$ for $1 \leq i \leq N$. Let $U^{1:N}=X^{1:N}G_N$. $\tilde{W}_N^{(i)}$ has input $U^i$ and outputs $(U^{1:i-1},Y^{1:N},H^{1:N})$. Since $\tilde{W}$ is symmetric, $\tilde{W}_N^{(i)}$ is symmetric as well \cite{arikan2009channel}. We can identify the information set according to the Bhattacharyya parameter $\tilde{Z}(\tilde{W}_N^{(i)})$. Treating $(Y,H)$ as the outputs, by Definition \ref{deft:symZ},
\begin{eqnarray}
\tilde{Z}(\tilde{W})=\sum\limits_{y,h} \sqrt{P_{Y,H|X}(y,h|+1)P_{Y,H|X}(y,h|-1)}.
\end{eqnarray}
Note that $\tilde{Z}(\tilde{W}_N^{(i)})$ can be evaluated recursively for BECs, starting with the initial Bhattacharyya parameter $\tilde{Z}(\tilde{W})$ (see \cite[eqn. (38)]{arikan2009channel}).
For general BMSCs, it is difficult to calculate $\tilde{Z}(\tilde{W}_N^{(i)})$ directly because of the exponentially increasing size of the output alphabet of $\tilde{W}_N^{(i)}$. Fortunately, we can apply the degrading and upgrading merging algorithms \cite{Ido, PolarConstru} to estimate $\tilde{Z}(\tilde{W}_N^{(i)})$ within acceptable accuracy.

In practice, the two approximations from the degrading and upgrading processes are rather close. Therefore, we focus on the degrading transform for brevity.

Define the likelihood ratio (LR) of $(y,h)$ as
\begin{eqnarray}
LR(y,h)\triangleq\frac{P_{Y,H|X}(y,h|+1)}{P_{Y,H|X}(y,h|-1)}.
\end{eqnarray}
By \eqref{eq:FadDens}, we have $LR(y,h)=e^{\frac{2yh}{\sigma^2}}$. Clearly, $LR(y,h) \geq 1$ for any $y \geq 0$. Each $LR(y,h)$ corresponds to a BSC with crossover probability $\frac{1}{LR(y,h)+1}$ and its capacity is given by
\begin{eqnarray}
C[LR(y,h)]=1-\mathfrak{h}_2\bigg (\frac{1}{LR(y,h)+1}\bigg),
\end{eqnarray}
where $\mathfrak{h}_2(\cdot)$ is the binary entropy function.

\begin{figure}[ht]
    \centering
    \includegraphics[width=9cm]{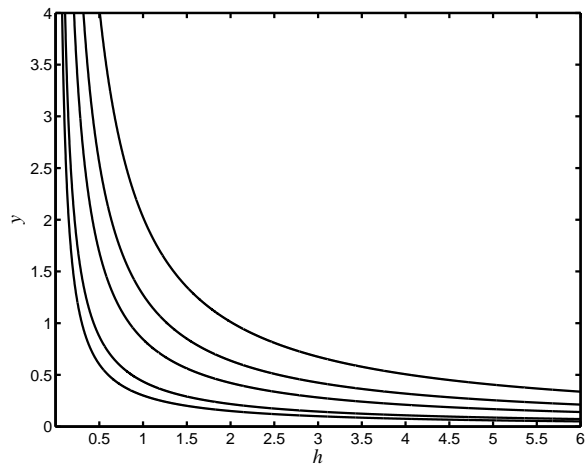}
    \caption{Typical boundaries of $A_i$ for channel quantization.}
    \label{fig:AiBoundary}
\end{figure}

The fading channel $\tilde{W}$ is then quantized according to $C[LR(y,h)]$. Let $\mu=2Q$ be the alphabet size of the degraded channel output alphabet. The set $\{y \geq 0, h \geq 0\}$ is divided into $Q$ subsets
\begin{eqnarray}
A_i=\bigg\{y \geq 0, h \geq 0: \frac{i-1}{Q} \leq C[LR(y,h)] < \frac{i}{Q}\bigg\},
\end{eqnarray}
for $1 \leq i \leq Q$. Typical boundaries of $A_i$ are depicted in Fig. \ref{fig:AiBoundary}. The outputs in $A_i$ are mapped to one symbol, and $\tilde{W}$ is quantized to a mixture of $Q$ BSCs with the crossover probability
\begin{eqnarray}
p_i=\frac{\int_{A_i}P_{Y,H|X}(y,h|-1) dydh}{\int_{A_i}P_{Y,H|X}(y,h|+1) dydh+\int_{A_i}P_{Y,H|X}(y,h|-1) dydh}.
\end{eqnarray}

Note that $p_i$ can be numerically evaluated. Since $LR(y,h)=e^{\frac{2yh}{\sigma^2}}$, $A_i$ is rewritten as
\begin{eqnarray}\label{eq:Ai}
A_i=\Bigg\{y \geq 0, h \geq 0: \frac{\sigma^2}{2} \ln\bigg(\frac{1}{\mathfrak{h}_2^{-1}(\frac{Q-i+1}{Q})}-1\bigg) \leq yh < \frac{\sigma^2}{2} \ln\bigg(\frac{1}{\mathfrak{h}_2^{-1}(\frac{Q-i}{Q})}-1\bigg) \Bigg\}.
\end{eqnarray}

Let $\delta_1$ and $\delta_2$ denote the two bounds in \eqref{eq:Ai}. We have
\begin{eqnarray}
\int_{A_i}P_{Y,H|X}(y,h|+1) dydh=\int_{0}^{\infty} \frac {h}{\sigma_h^2} e^{-\frac{h^2}{2\sigma_h^2}}dh \int_{\frac{\delta_1}{h}}^{\frac{\delta_2}{h}} \frac{1}{\sqrt{2 \pi \sigma^2}} e^{-\frac{(y-h)^2}{2\sigma^2}} dy,
\end{eqnarray}
and $\int_{A_i}P_{Y,H|X}(y,h|-1) dydh$ is calculated similarly.

\begin{figure}[ht]
    \centering
    \includegraphics[width=9cm]{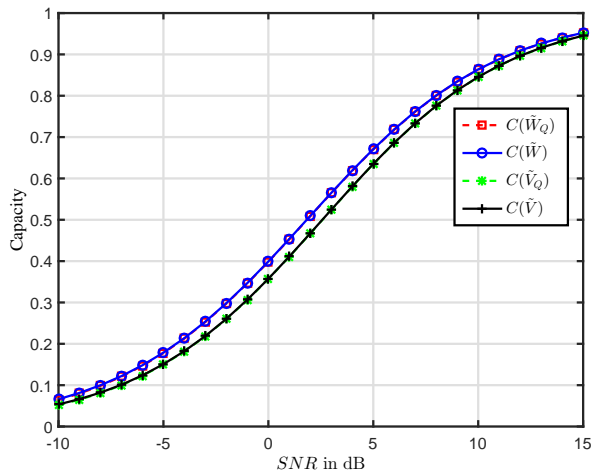}
    \caption{Comparisons between $C(\tilde{W}_Q)$ and $C(\tilde{W})$, and between $C(\tilde{V}_Q)$ and $C(\tilde{V})$, when $Q=128$. Here $\tilde{W}$ denotes the channel $X \to (Y,H)$, and $\tilde{V}$ denotes the channel $X \to Y$, i.e., the two channel models when the receiver knows the CSI and the CDI, respectively. $\tilde{W}_Q$ and $\tilde{V}_Q$ denote the quantized version of $\tilde{W}$ and $\tilde{V}$, respectively.}
    \label{fig:CDiff}
\end{figure}

Let $\tilde{W}_Q$ denote the quantized channel from $\tilde{W}$ after the degrading transform. By \cite[Lemma 13]{Ido}, the difference between the two channel capacities is upper-bounded by $\frac{1}{Q}$. A comparison between $C(\tilde{W}_Q)$ and $C(\tilde{W})$ for different $SNR$ when $Q=128$ is shown in Fig. \ref{fig:CDiff}. When $Q$ is sufficiently large, we can use $\tilde{W}_Q$ to approximate $\tilde{W}$ in the construction of polar codes. The size of the output alphabet after the degrading merging is no more than $2Q$.

The proof of the following theorem can be adapted from \cite{Ido}. We omit it for brevity.

\begin{ther}
Let $\tilde{W}: X \to (Y,H)$ be a binary-input i.i.d. fading channel. Let $N$ denote the block length and $\mu=2Q$ denote the limit of the size of output alphabet. A polar code constructed by the degrading merging algorithm achieves the capacity $C(\tilde{W})$ when $N$ and $\mu$ are both sufficiently large. The block error probability under SC decoding is upper-bounded by $N2^{-N^\beta}$ for $0<\beta<\frac{1}{2}$.
\end{ther}

\begin{figure}[ht]
    \centering
    \includegraphics[width=9cm]{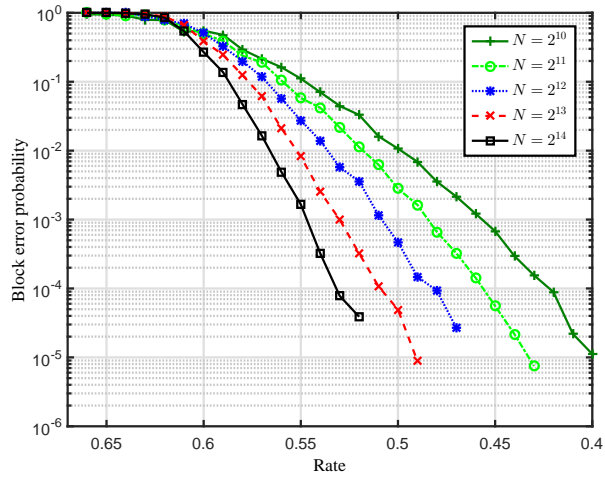}
    \caption{Performance of polar codes for the Rayleigh fading channel with CSI available to the receiver when $N=2^{10},2^{11},...,2^{14}$.}
    \label{fig:bfadsimu}
\end{figure}

Simulation result of polar codes with different block length for the binary-input Rayleigh fading channel with CSI available to the receiver are shown in Fig. \ref{fig:bfadsimu}, where $SNR=5$ dB and $C(\tilde{W})=0.6709$. The performance can be further improved by using more sophisticated decoding algorithms \cite{ListPolar,eslami2012finite}.

\begin{figure}[ht]
    \centering
    \includegraphics[width=9cm]{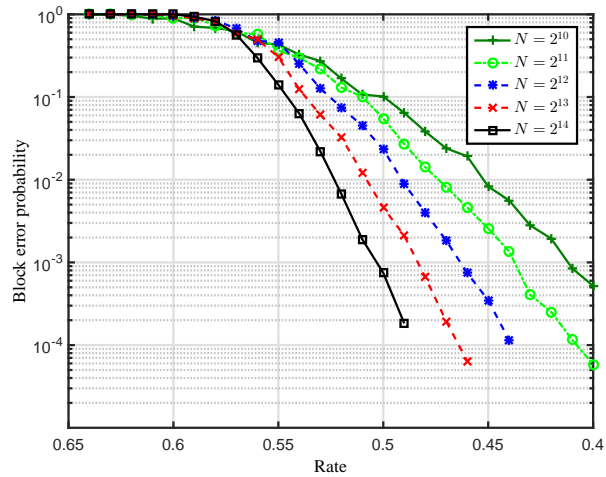}
    \caption{Performance of polar codes for the Rayleigh fading channel with CDI available to the receiver when $N=2^{10},2^{11},...,2^{14}$.}
    \label{fig:bfadsimu_CDI}
\end{figure}

\begin{rem}
It has been pointed out in \cite{PolarFadingAngel} that polar codes for the Rayleigh fading channel with known CDI suffer a penalty for not having complete information. The statement can be seen clearly from our construction. Treating $H$ as part of channel outputs, the binary channel $X \to Y$ is degraded with respect to the channel $X \to (Y,H)$, and $I(X;Y,H) \geq I(X;Y)$. Let $\tilde{V}$ denote the channel $X \to Y$. The channel transition PDF of $\tilde{V}$ is written as
\begin{eqnarray}
P_{Y|X}(y|x)=\int_{h} P_{Y,H|X}(y,h|x) dh,
\end{eqnarray}
where $P_{Y,H|X}(y,h|x)$ is given by \eqref{eq:FadDens}. It is clear that $\tilde{V}$ is a BMSC. Therefore, the degrading and upgrading merging algorithms can also be applied to construct polar codes for $\tilde{V}$. A comparison between $C(\tilde{W})$ and $C(\tilde{V})$ is shown in Fig. \ref{fig:CDiff}. By Remark \ref{rk:polardegrade},
the polar code constructed when the receiver only knows the CDI is a subcode of that when the receiver knows the CSI. Simulation results of polar codes for the binary-input Rayleigh fading channel with CDI available to the receiver are shown in Fig. \ref{fig:bfadsimu_CDI}, where $SNR=5$ dB and $C(\tilde{V})=0.6352$.
\end{rem}

\begin{figure}[ht]
    \centering
    \includegraphics[width=9cm]{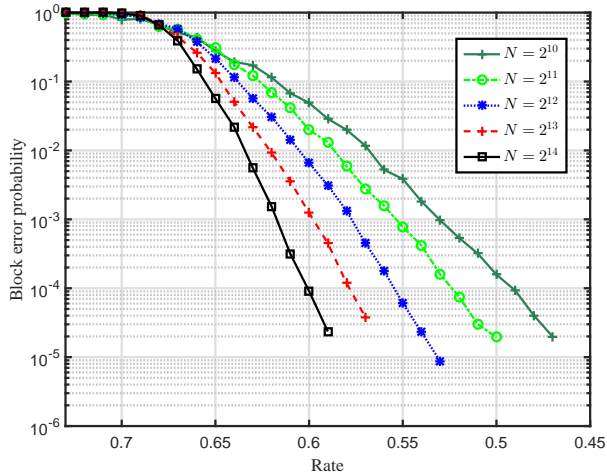}
    \caption{Performance of polar codes for the Rician fading channel with CSI available to the receiver when $N=2^{10},2^{11},...,2^{14}$.}
    \label{fig:bfadsimu_rician}
\end{figure}

\begin{rem}
Our construction method can be generalized to other fading distributions such as the Rician distribution, the lognormal distribution and the Nakagami distribution. Taking the Rician distribution as an example, the PDF of $H$ becomes
\begin{eqnarray}\label{eq:rician}
P_H(h)=\frac{h}{\sigma_h^2} e^{-\frac{(h^2+s^2)}{2\sigma_h^2}}I_0\Big(\frac{hs}{\sigma_h^2}\Big),
\end{eqnarray}
where $\sigma_h$ is the scale parameter, $s$ is the non-centrality parameter, and $I_0(\cdot)$ is the modified Bessel function of the first kind with order zero. $P_{Y,H|X}(y,h|x)$, $C(\tilde{W})$ and $LR(y,h)$ can be calculated similarly, with $P_H(h)$ being replaced by \eqref{eq:rician}. We apply the same channel quantization method to construct polar codes. Performance of polar codes for the binary-input Rician fading channel with CSI available to the receiver is shown in Fig. \ref{fig:bfadsimu_rician}, where $\frac{\sigma_h^2}{\sigma^2}=5$ dB, $s=1$, and $C(\tilde{W})=0.7326$.
\end{rem}

\section{Polar Lattices for i.i.d. Fading Channels}\label{sec:GFading}
In this section, we extend polar codes to polar lattices for i.i.d. fading channels. The reason for this extension is that the input of fading channels is not necessarily limited to be binary. In general, the input $X$ is subject to a power constraint $P$, i.e.,
\begin{eqnarray}
E[X^2]\leq P.
\end{eqnarray}

In this case, lattice codes offer more choices of input constellation. It has been shown in \cite{forney6} that lattice codes are able to achieve the sphere bound, or the Poltyrev capacity of AWGN channels. These codes are defined as AWGN-good lattices. To achieve the AWGN capacity, the AWGN-good lattices should be properly shaped to obtain the optimum shaping gain. This can be accomplished by using lattices which are good for quantization \cite{zamir} or by the lattice Gaussian shaping technique \cite{LingBel13}. An explicit construction of the AWGN-good polar lattices with lattice Gaussian shaping was presented in \cite{polarlatticeJ}. Our work follows a similar line. We firstly construct polar lattices which achieve the Poltyrev capacity of i.i.d. fading channels and then perform lattice Gaussian shaping to achieve the ergodic capacity. Before that, we give a brief review of the construction of the AWGN-good polar lattices.

\subsection{AWGN-Good Polar Lattices}\label{sec:Agood}
A mod-$\Lambda$ Gaussian channel is a Gaussian channel with an input in $\mathcal{V}(\Lambda)$ and with a mod-$\mathcal{V}(\Lambda)$ operator at the receiver front end \cite{forney6}. The capacity of the mod-$\Lambda$ channel with noise variance $\sigma^2$ is
\begin{eqnarray}
C(\Lambda, \sigma^{2})=\log  V(\Lambda)-\mathfrak{h}(\Lambda, \sigma^{2}),
\end{eqnarray}
where $\mathfrak{h}(\Lambda,\sigma^{2})=-\int_{\mathcal{V}(\Lambda)}f_{\sigma,\Lambda}({x})\text{ log } f_{\sigma,\Lambda}({x})d{x}$ is the differential entropy of the $\Lambda$-aliased noise over $\mathcal{V}(\Lambda)$.

\begin{rem}\label{rk:modLambda}
A mod-$\Lambda$ Gaussian channel with noise variance $\sigma_1^2$ is degraded with respect to one with noise variance $\sigma_2^2$ if $\sigma_1^2 > \sigma_2^2$. Let $\tilde{W}_1$ and $\tilde{W}_2$ denote the two channels respectively. Consider an intermediate channel $\tilde{W}'$ which is also a mod-$\Lambda$ channel, with noise variance $\sigma_1^2-\sigma_2^2$. By the property $[X \mod \Lambda + Y] \mod \Lambda= [X+Y] \mod \Lambda$, it is easy to see that $\tilde{W}_1$ is stochastically equivalent to a channel constructed by concatenating $\tilde{W}_2$ with $\tilde{W}'$. Therefore, $C(\Lambda, \sigma_1^{2}) < C(\Lambda, \sigma_2^{2})$, and $\mathfrak{h}(\Lambda, \sigma_1^{2}) > \mathfrak{h}(\Lambda, \sigma_2^{2})$.
\end{rem}

A sublattice $\Lambda' \subset \Lambda$ induces a partition (denoted by $\Lambda/\Lambda'$) of $\Lambda$ into equivalence classes modulo $\Lambda'$. For a lattice partition $\Lambda/\Lambda'$, the $\Lambda/\Lambda'$ channel is a mod-$\Lambda'$ channel whose input is restricted to discrete lattice points in $(\Lambda + a) \cap \mathcal{R}(\Lambda')$ for some translate $a$. The order of the partition is denoted by $|\Lambda/\Lambda'|$, which is equal to the number of cosets. If $|\Lambda/\Lambda'|=2$, we call this a binary partition. The capacity of the $\Lambda/\Lambda'$ channel is given by \cite{forney6}
\begin{equation}\label{mod12-capacity}
\begin{split}
  C(\Lambda/\Lambda', \sigma^2) &= C(\Lambda', \sigma^2) - C(\Lambda, \sigma^2) \\
  &= \mathfrak{h}(\Lambda, \sigma^2) - \mathfrak{h}(\Lambda', \sigma^2) + \log \big(V(\Lambda')/V(\Lambda)\big).
\end{split}
\end{equation}

\begin{rem}\label{rk:modLambda'}
The $\Lambda/\Lambda'$ channel is symmetric \cite{forney6}. Similar to Remark \ref{rk:modLambda}, a $\Lambda/\Lambda'$ channel with noise variance $\sigma_1^2$ is degraded with respect to one with noise variance $\sigma_2^2$, if $\sigma_1^2 > \sigma_2^2$. Therefore, $C(\Lambda/\Lambda', \sigma_1^2) < C(\Lambda/\Lambda', \sigma_2^2)$. Moreover, for a self-similar partition $\Lambda_0/\Lambda_1/\Lambda_2$ and a fixed noise variance $\sigma^2$, the $\Lambda_1/\Lambda_2$ channel at higher level is stochastically equivalent with a $\Lambda_0/\Lambda_1$ channel with smaller noise variance than $\sigma^2$. Therefore, the $\Lambda_0/\Lambda_1$ channel is degraded with respect to the $\Lambda_1/\Lambda_2$ channel, and $C(\Lambda_0/\Lambda_1, \sigma^2) < C(\Lambda_1/\Lambda_2, \sigma^2)$. See the proof in \cite{polarlatticeJ} for more details.
\end{rem}

As we mentioned, we use the ``Construction D" method to construct polar lattices. Let $\Lambda/\Lambda_{1}/\cdots/\Lambda_{r-1}/\Lambda'$ for $r \geq 1$ be an $n$-dimensional self-similar lattice partition chain. For each
partition $\Lambda_{\ell-1}/\Lambda_{\ell}$ ($1\leq \ell \leq r$ with convention $\Lambda_0=\Lambda$ and $\Lambda_r=\Lambda'$) a code over $\Lambda_{\ell-1}/\Lambda_{\ell}$
selects a sequence of representatives $a_{\ell}$ for the cosets of $\Lambda_{\ell}$. If each partition is a binary partition, the codes $\mathcal{C}_{\ell}$ are binary codes. Moreover, based on this partition chain, the capacity $C(\Lambda/\Lambda', \sigma^2)$ can be expanded as
\begin{equation}
C(\Lambda/\Lambda', \sigma^2) = C(\Lambda/\Lambda_1, \sigma^2) + \cdots + C(\Lambda_{r-1}/\Lambda', \sigma^2).
\end{equation}

The key idea of the AWGN-good polar lattices is to use a good component polar code to achieve the capacity $C(\Lambda_{\ell-1}/\Lambda_{\ell}, \sigma^2)$ for each level $\ell=1,2,\ldots,r$. A polar lattice $L$ is resulted from those component polar codes. For such a construction, the total decoding error probability with multi-stage decoding is bounded by
\begin{equation}\label{total-Pe}
    P_e(L, \sigma^{2}) \leq \sum_{\ell=1}^{r}{ P_e(\mathcal{C}_{\ell},\sigma^{2})} + P_e\big((\Lambda')^{N},\sigma^{2}\big),
\end{equation}
where $P_e(\mathcal{C}_{\ell},\sigma^{2})$ denotes the decoding error probability of polar code $\mathcal{C}_{\ell}$ at level $\ell$. To make $P_e(L, \sigma^{2}) \to 0$, we need to choose the bottom lattice $\Lambda'$ such that the uncoded error probability $P_e\big((\Lambda')^{N},\sigma^{2}\big) \to 0$ and construct a code $\mathcal{C}_{\ell}$ for each $\Lambda_{\ell-1}/\Lambda_{\ell}$ channel such that the decoding error probability $P_e(\mathcal{C}_{\ell},\sigma^{2})$ also tends to zero. Note that the mod-$\Lambda$ channel is not used for communication and $C(\Lambda, \sigma^{2})$ is required to be negligible.

To sum up, in order to approach the Poltyrev capacity of AWGN channels, we would like to have $\log\Big(\frac{\gamma_{L}(\sigma)}{2\pi e}\Big) \to 0$ while $P_e(L, \sigma^2) \to 0$. According to the analysis in \cite{forney6}, we have the following three design criteria:
\begin{itemize}
  \item The top lattice $\Lambda$ gives negligible capacity $C(\Lambda, \sigma^2)$.
  \item The bottom lattice $\Lambda'$ has a small error probability $P_e(\Lambda',\sigma^{2})$.
  \item Each component polar code $\mathcal{C}_{\ell}$ is a capacity-approaching code for the $\Lambda_{\ell-1} / \Lambda_{\ell}$ channel.
\end{itemize}
Since polar codes are capacity-achieving, polar lattices are proved to be AWGN-good for a properly chosen lattice partition \cite{polarlatticeJ}. The concepts of mod-$\Lambda$ channel, $\Lambda/\Lambda'$ channel and AWGN-goodness will be generalized to fading channels in the next subsection.

\subsection{Polar Lattices for Fading channels Without Power Constraint}\label{sec:notPower}
For i.i.d. fading channels, the channel gain varies. The above analysis for AWGN channels need to be generalized. Since the receiver knows the CSI, the fading effect can be removed by multiplying $Y$ with $\frac{1}{H}$. We define the fading mod-$\Lambda$ channel as follows.

\begin{figure}[ht]
    \centering
    \includegraphics[width=10cm]{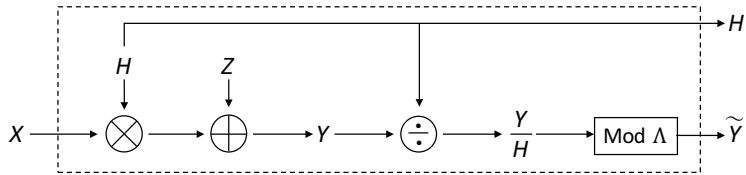}
    \caption{A block diagram of the fading mod-$\Lambda$ channel.}
    \label{fig:modfading}
\end{figure}

\begin{definition}[fading mod-$\Lambda$ channel]
A fading mod-$\Lambda$ channel is a fading channel with an input in $\mathcal{V}(\Lambda)$, and an output being scaled by $\frac{1}{H}$ before the mod-$\mathcal{V}(\Lambda)$ operation. A block diagram of this model is shown in Fig. \ref{fig:modfading}.
\end{definition}

Note that here we assume that the fading coefficient remains the same during $n$ transmission symbols, where $n$ is the dimension of lattice $\Lambda$. A proper $n$ can be chosen according to the coherence time of the fading channel. The fading coefficient is also assumed to be independent between different blocks. The fading mod-$\Lambda$ channel is closely related to a mod-$\Lambda$ channel with noise variance $\frac{\sigma^2}{h^2}$. For convenience, letting $n=1$, the channel transition PDF of the fading mod-$\Lambda$ channel is given by
\begin{equation}\label{eq:modFadingPDF}
\begin{aligned}
P_{\tilde{Y},H|X}(\tilde{y},h|x)&=P_{Y,H|X}(y=\tilde{y}h+h\cdot \Lambda,h|x)\frac{dy}{d\tilde{y}} \\
&=h\cdot P_H(h) \sum_{\lambda \in \Lambda} P_{Y|X,H}(y=\tilde{y}h+\lambda h|x,h) \\
&=h\cdot P_H(h) \sum_{\lambda \in \Lambda} \frac{1}{\sqrt{2 \pi} \sigma} e^{-\frac{(\tilde{y}h+\lambda h-xh)^2}{2\sigma^2}}\\
&=P_H(h) \sum_{\lambda \in \Lambda} \frac{1}{\sqrt{2 \pi} \frac{\sigma}{h}}e^{-\frac{(\tilde{y}+\lambda-x)^2}{2\big(\frac{\sigma}{h}\big)^2}},
\end{aligned}
\end{equation}
where the second term in the last equation is the channel transition PDF of a mod-$\Lambda$ channel with noise variance $\frac{\sigma^2}{h^2}$. The channel transition PDF for higher dimension $n$ can be derived similarly. As a result, the fading mod-$\Lambda$ channel can be viewed as an independent combination of a Rayleigh distributed variable $H$ and a mod-$\Lambda$ channel with noise variance $\frac{\sigma^2}{H^2}$. The capacity of the fading mod-$\Lambda$ channel is
\begin{equation}\label{eq:CH}
\begin{aligned}
C_H(\Lambda,\sigma^2)&=C(X;\tilde{Y},H)\\
&=C(X;\tilde{Y}|H) \\
&=\int_h P_H(h) C(X;\tilde{Y}|h) dh \\
&=E_h\bigg[C\bigg(\Lambda,\frac{\sigma^2}{h^2}\bigg)\bigg]\\
&=\log V(\Lambda)-E_h\bigg[\mathfrak{h}\bigg(\Lambda, \frac{\sigma^2}{h^2}\bigg)\bigg].
\end{aligned}
\end{equation}

Similarly, a fading $\Lambda/\Lambda'$ channel is a fading mod-$\Lambda'$ channel whose input is restricted to discrete lattice points in $(\Lambda + a) \cap \mathcal{R}(\Lambda')$ for some translate $a$. By the same argument of \eqref{eq:modFadingPDF}, it can be viewed as an independent combination of a Rayleigh distributed variable $H$ and a $\Lambda/\Lambda'$ channel with noise variance $\frac{\sigma^2}{H^2}$. The capacity of the fading $\Lambda/\Lambda'$ channel is given by
\begin{equation}
\begin{aligned}
C_H(\Lambda/\Lambda',\sigma^2)&=E_h\bigg[C\bigg(\Lambda',\frac{\sigma^2}{h^2}\bigg)\bigg]-E_h\bigg[C\bigg(\Lambda,\frac{\sigma^2}{h^2}\bigg)\bigg]\\
&=E_h\bigg[\mathfrak{h}\bigg(\Lambda, \frac{\sigma^2}{h^2}\bigg)\bigg] - E_h\bigg[\mathfrak{h}\bigg(\Lambda', \frac{\sigma^2}{h^2}\bigg)\bigg] + \log \big(V(\Lambda')/V(\Lambda)\big).
\end{aligned}
\end{equation}

For a self-similar partition chain $\Lambda/\Lambda_{1}/\cdots/\Lambda_{r-1}/\Lambda'$, we have
\begin{equation}
C_H(\Lambda/\Lambda', \sigma^2) = C_H(\Lambda/\Lambda_1, \sigma^2) + \cdots + C_H(\Lambda_{r-1}/\Lambda', \sigma^2).
\end{equation}

Since the $\Lambda/\Lambda'$ channel is symmetric, it is easy to check that the $\Lambda/\Lambda'$ fading channel is symmetric as well. Moreover, if $|\Lambda/\Lambda'|=2$, the $\Lambda/\Lambda'$ fading channel is a BMSC. Taking the binary partition $\mathbb{Z}/2\mathbb{Z}$ as an example, the input of the $\mathbb{Z}/2\mathbb{Z}$ fading channel is $X \in \{0,1\}$, and a permutation $\phi$ over the outputs $(\tilde{y},h)$ is defined such that $\phi(\tilde{y},h)=([\tilde{y}-1] \mod 2\mathbb{Z},h)$. Check that $P_{\tilde{Y},H|X}(\tilde{y},h|0)=P_{\tilde{Y},H|X}(\phi(\tilde{y},h)|1)$.

It is now clear that polar lattices can be constructed to achieve the (ergodic) Poltyrev capacity of the i.i.d. fading channels, as we did for the AWGN channel in Sect. \ref{sec:Agood}. Recall that the Poltyrev capacity $C_{\infty}$ of a
general additive-noise channel is defined as the capacity per unit volume in \cite[Theorem 6.3.1]{BK:Zamir}. For the independent AWGN channel, we have
\begin{equation}
C_{\infty}=-\mathfrak{h}(\sigma^2)=\frac{1}{2}\log\bigg(\frac{1}{2 \pi e \sigma^2 }\bigg),
\end{equation}
where $\mathfrak{h}(\sigma^2)$ denotes the differential entropy of a Gaussian random variable with variance $\sigma^2$.

For independent fading channels, $C_{\infty}$ is generalized as \cite{ShlomiFading}
\begin{equation}
C_{\infty}=-E_h\bigg[\mathfrak{h}\bigg(\frac{\sigma^2}{h^2}\bigg)\bigg]=E_h\bigg[\frac{1}{2}\log \bigg(\frac{h^2}{2 \pi e \sigma^2}\bigg)\bigg].
\end{equation}

In the special case of Rayleigh fading,
\begin{equation}
\begin{aligned}
C_{\infty}&=-\int_{h} \frac{h}{\sigma_h^2} e^{-\frac{h^2}{2\sigma_h^2}} \frac{1}{2}\log\bigg(\frac{2 \pi e \sigma^2}{h^2}\bigg) dh \\
&\hspace{-.75em}\underset{t=\frac{h^2}{2\sigma_h^2}}{=}-\frac{1}{2} \int_{t} e^{-t} \bigg(\log \bigg(\frac{2 \pi e \sigma^2}{2 \sigma_h^2}\bigg)- \log t\bigg) dt\\
&=-\frac{1}{2} \log\bigg (2\pi e \sigma^2 \cdot \frac{e^{\zeta}}{2\sigma_h^2}\bigg),
\end{aligned}
\end{equation}
where $\zeta=-\int_{0}^{\infty} e^{-x} \ln x dx$ is the Euler-Mascheroni constant.

To approach the Poltyrev capacity $-\frac{1}{2} \log \Big(2\pi e \sigma^2 \cdot \frac{e^{\zeta}}{2\sigma_h^2}\Big)$, we construct polar lattices according to the following three design criteria:
\begin{itemize}
  \item[(a)] The top lattice $\Lambda$ gives negligible capacity $E_h\Big[C\Big(\Lambda, \frac{\sigma^2}{h^2}\Big)\Big]$.
  \item[(b)] The bottom lattice $\Lambda'$ has a small error probability $E_h\Big[P_e\Big(\Lambda',\frac{\sigma^2}{h^2}\Big)\Big]$.
  \item[(c)] Each component polar code $\mathcal{C}_{\ell}$ is a capacity-approaching code for the $\Lambda_{\ell-1} / \Lambda_{\ell}$ fading channel.
\end{itemize}

For criterion (a), we pick a top lattice $\Lambda$ for a large channel gain $h_l$ such that $\mathfrak{h}\Big(\Lambda, \frac{\sigma^2}{h_l^2}\Big) \approx \log V(\Lambda)$. By Remark \ref{rk:modLambda}, $\mathfrak{h}\Big(\Lambda, \frac{\sigma^2}{h^2}\Big) \geq \mathfrak{h}\Big(\Lambda, \frac{\sigma^2}{h_l^2}\Big)$ for $0 \leq h \leq h_l$.
\begin{equation}\label{eq:EHapprox}
\begin{aligned}
E_h\bigg[\mathfrak{h}\bigg(\Lambda, \frac{\sigma^2}{h^2}\bigg)\bigg]&= \int_{0}^{h_l} P_H(h) \mathfrak{h}\bigg(\Lambda, \frac{\sigma^2}{h^2}\bigg) dh +\int_{h_l}^{\infty} P_H(h) \mathfrak{h}\bigg(\Lambda, \frac{\sigma^2}{h^2}\bigg) dh \\
&\gtrapprox  \mathfrak{h}\bigg(\Lambda, \frac{\sigma^2}{h_l^2}\bigg)\int_{0}^{h_l} P_H(h) dh + n\int_{h_l}^{\infty} P_H(h) \mathfrak{h}\bigg(\frac{\sigma^2}{h^2}\bigg) dh \\
&=\mathfrak{h}\bigg(\Lambda,\frac{\sigma^2}{h_l^2}\bigg)\bigg(1-e^{-\frac{h_l^2}{2\sigma_h^2}}\bigg)+\frac{n}{2} \log \bigg(\frac{2 \pi e \sigma^2}{h_l^2}\bigg)e^{-\frac{h_l^2}{2\sigma_h^2}} -\frac{n}{2} \log e \cdot E_1\bigg(\frac{h_l^2}{2\sigma_h^2}\bigg),
\end{aligned}
\end{equation}
where $E_1(x)=\int_{x}^{\infty} \frac{e^{-t}}{t} dt$ is the exponential integral, and $E_1(x) \to 0$ for $x \to \infty$. The approximation is due to the fact $\mathfrak{h}\Big(\Lambda, \frac{\sigma^2}{h^2}\Big) \to n\mathfrak{h}\Big(\frac{\sigma^2}{h^2}\Big)$ as $h \to \infty$. Let $h_l=O(N)$. We have $E_h\Big[\mathfrak{h}\Big(\Lambda, \frac{\sigma^2}{h^2}\Big)\Big] \approx \log V(\Lambda)$, and $E_h\Big[C\Big(\Lambda, \frac{\sigma^2}{h^2}\Big)\Big]\approx 0$ as $N \to \infty$ according to \eqref{eq:CH}.

For criterion (b), we pick a bottom lattice $\Lambda'$ for a small channel gain $h_s$ such that $P_e\big(\Lambda', \frac{\sigma^2}{h_s^2}\big) \to 0$. Since $P_e\big(\Lambda', \frac{\sigma^2}{h^2}\big) \leq P_e\big(\Lambda', \frac{\sigma^2}{h_s^2}\big)$ for $h \geq h_s$,
\begin{equation}\label{eq:EPe}
\begin{aligned}
E_h\bigg[P_e\bigg(\Lambda', \frac{\sigma^2}{h^2}\bigg)\bigg]&=\int_{0}^{h_s} P_H(h) P_e\bigg(\Lambda', \frac{\sigma^2}{h^2}\bigg) dh+\int_{h_s}^{\infty} P_H(h) P_e\bigg(\Lambda', \frac{\sigma^2}{h^2}\bigg) dh \\
&\leq \bigg(1-e^{-\frac{h_s^2}{2 \sigma_h^2}}\bigg)+ P_e\bigg(\Lambda', \frac{\sigma^2}{h_s^2}\bigg) \cdot e^{-\frac{h_s^2}{2\sigma_h^2}}.
\end{aligned}
\end{equation}
Let $h_s=O\big(\frac{1}{N^\delta}\big)$ for some constant $\delta \geq 1$. We have $E_h\Big[P_e\Big(\Lambda',\frac{\sigma^2}{h^2}\Big)\Big] \to 0$ as $N \to \infty$. Since the volume $V(\Lambda')$ is sufficiently large to cover almost all of the noised signal, by \cite{forney6}, we have $E_h\Big[\mathfrak{h}\Big(\Lambda' ,\frac{\sigma^2}{h^2}\Big)\Big] \approx nE_h\Big[\mathfrak{h}\Big(\frac{\sigma^2}{h^2}\Big)\Big]$ when $E_h \Big[P_e\Big(\Lambda',\frac{\sigma^2}{h^2}\Big)\Big] \to 0$. Note that $\delta$ is required to be lager than 1 here to guarantee that $E_h \Big[P_e\Big(\Lambda',\frac{\sigma^2}{h^2}\Big)\Big]$ vanishes polynomially (see the proof of Theorem \ref{thm:codinggood}).

\begin{figure}[ht]
    \centering
    \includegraphics[width=15cm]{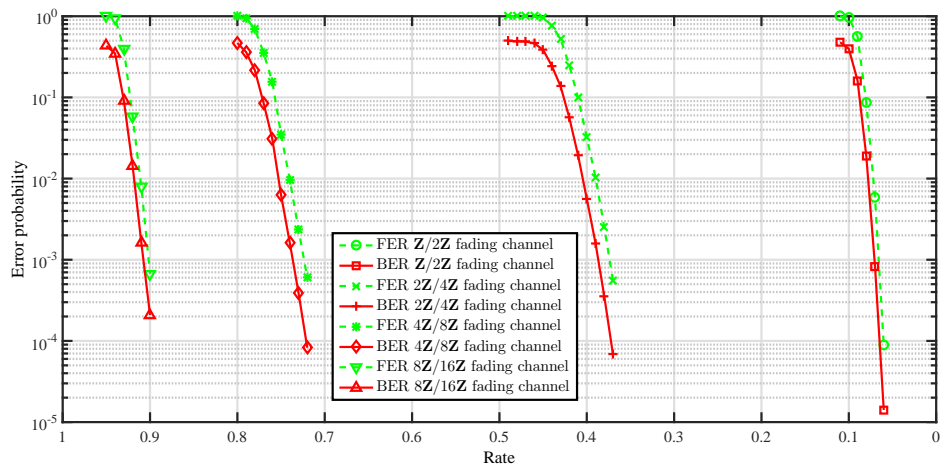}
    \caption{Performance of polar codes for the $\mathbb{Z}/2\mathbb{Z}$, $2\mathbb{Z}/4\mathbb{Z}$, $4\mathbb{Z}/8\mathbb{Z}$ and $8\mathbb{Z}/16\mathbb{Z}$ fading channels with $\sigma=1$, $\sigma_h=1.2575$ and $N=2^{14}$. The capacities of these four channels are about 0.1172, 0.4929, 0.8200 and 0.9500, respectively. FER denotes the frame (block) error probability, and BER denotes the bit error probability.}
    \label{fig:modsimu}
\end{figure}

For criterion (c), we choose a binary partition chain and construct binary polar codes to achieve the capacity of the $\Lambda_{\ell-1} / \Lambda_{\ell}$ fading channel for $1 \leq \ell \leq r$. Since the $\Lambda_{\ell-1} / \Lambda_{\ell}$ fading channel is a BMSC, treating $(\tilde{Y},H)$ as the outputs, the construction method proposed in Sect. \ref{sec:bfading} can be used. It remains to verify $C_{\ell-1} \subseteq C_\ell$. Since the the $\Lambda/\Lambda'$ fading channel can be viewed as an independent combination of a Rayleigh distributed variable $H$ and a $\Lambda/\Lambda'$ channel with noise variance $\frac{\sigma^2}{H^2}$, by Remark \ref{rk:modLambda'} and Remark \ref{rk:polardegrade}, we immediately have $C_{\ell-1} \subseteq C_\ell$. Simulation results of polar codes for the one-dimensional binary partition chain $\mathbb{Z}/2\mathbb{Z}/4\mathbb{Z}/8\mathbb{Z}/16\mathbb{Z}$ with $\sigma=1$, $\sigma_h=1.2575$ and block length $N=2^{14}$ are shown in Fig. \ref{fig:modsimu}.

\begin{ther}
[Good polar lattices for fading channel]\label{thm:codinggood}
For an independent Rayleigh fading channel with given $\sigma_h^2$ and $\sigma^2$, select an $n$-dimensional binary lattice partition chain $\Lambda/\Lambda_1/\cdots/\Lambda_{r-1}/\Lambda'$ such that both the criterion (a) and (b) are satisfied. Construct a polar lattice $L$ from this partition chain and $r$ nested polar codes with block length $N$. Let $r=n\delta O(\log N)$ for a fixed dimension $n$ and some constant $\delta \geq 1$. $L$ can achieve the Poltyrev capacity of the i.i.d. fading channel, i.e.,
$\gamma_{L}(\sigma)\rightarrow 2 \pi e \cdot \frac{e^\zeta}{2\sigma_h^2}$
and
$P_{e}(L, \sigma^{2})=O\big(\frac{1}{N^{2\delta-1}}\big)\rightarrow0$, as $N\to \infty$.
\end{ther}

\begin{IEEEproof}
By the the union bound of the error probability under the multi-stage lattice decoding \cite{forney6}, $P_{e}(L, \sigma^{2})$ is upper-bounded by
\begin{eqnarray}
P_{e}(L, \sigma^{2})\leq rN2^{-N^{\beta}}+N \cdot E_h\bigg[P_e\bigg(\Lambda',\frac{\sigma^2}{h^2}\bigg)\bigg].
\label{eqn:errorbound}
\end{eqnarray}

Let $h_s=O\big(\frac{1}{N^\delta}\big)$ for some constant $\delta \geq 1$ be a small channel gain and let $h_l=O(N)$ be a large channel gain. Consider a fine lattice $\Lambda_f$ and a coarse lattice $\Lambda_c$ in the lattice partition chain such that $\mathfrak{h}(\Lambda_f,\sigma^2) \approx \log V(\Lambda_f)$ and $P_e(\Lambda_c,\sigma^2) \to 0$. Let $d$ be the minimum distance of $\Lambda_c$. By the Chernoff bound, we have
\begin{eqnarray}
P_e(\Lambda_c,\sigma^2)\leq nQ\bigg(\frac{d}{2\sigma}\bigg) \leq n \exp\bigg(-\frac{d^2}{8\sigma^2}\bigg),
\end{eqnarray}
when $Q(\cdot)$ denotes the Q-function. Let $d=O(\sqrt{N})$ for a fixed $n$, $P_e(\Lambda_c,\sigma^2)$ decays exponentially. In this case, the number of partition levels between $\Lambda_f$ and $\Lambda_c$ is $nO(\log N)$. We further let $\Lambda=\frac{1}{h_l} \Lambda_f$ and $\Lambda'=\frac{1}{h_s} \Lambda_c$. Check that $\mathfrak{h}(\Lambda_f, \sigma^2)=\mathfrak{h}\Big(\Lambda, \frac{\sigma^2}{h_l^2}\Big)+\log \big(V(\Lambda_f)/V(\Lambda)\big)$ and $P_e\Big(\Lambda',\frac{\sigma^2}{h_s^2}\Big)=P_e(\Lambda_c,\sigma^2)$, which means $\mathfrak{h}\Big(\Lambda, \frac{\sigma^2}{h_l^2}\Big) \approx \log V(\Lambda)$ and $P_e\Big(\Lambda',\frac{\sigma^2}{h_s^2}\Big)=e^{-O(N)}$. Therefore, criteria (a) and (b) are satisfied when $N \to \infty$. The number $r$ of levels between $\Lambda$ and $\Lambda'$ is given by
\begin{eqnarray}\label{eq:levelr}
\begin{aligned}
r&=\log \big( V(\Lambda')/V(\Lambda)\big) \\
&= \log \big( V(\Lambda_f)/V(\Lambda)\big) + \log \big( V(\Lambda_c)/V(\Lambda_f)\big) +\log \big( V(\Lambda')/V(\Lambda_c)\big) \\
&= n \log (h_l/h_s) + \log \big( V(\Lambda_c)/V(\Lambda_f)\big) \\
&=n\delta O(\log N).
\end{aligned}
\end{eqnarray}
Moreover, according to \eqref{eq:EPe}, $E_h\big[P_e\big(\Lambda', \frac{\sigma^2}{h^2}\big)\big]=O\big(\frac{1}{N^{2\delta}}\big)$, and then $P_e(L,\sigma^2)=O\big(\frac{1}{N^{2\delta-1}}\big)$.

Let $R_{\mathcal{C}}=\sum_{\ell=1}^{r} R_{\ell}$ be the total rate of polar codes from level 1 to level $r$. Since $V(L)=2^{-NR_{\mathcal{C}}}V(\Lambda')^{N}$, the logarithmic VNR of $L$ is
{\allowdisplaybreaks\begin{eqnarray}
\log \left(\frac{\gamma_{L}(\sigma)}{2\pi e} \cdot \frac{2\sigma_h^2}{e^\zeta}\right)
&=&\log \left( \frac{V(L)^\frac{2}{nN}}{2\pi e \sigma^2} \cdot \frac{2\sigma_h^2}{e^\zeta} \right) \\
&=&\log \left( \frac{2^{-\frac{2}{n}R_{\mathcal{C}}}V(\Lambda')^{\frac{2}{n}}}{2\pi e \sigma^2} \cdot \frac{2\sigma_h^2}{e^\zeta}\right)\\
&=&-\frac{2}{n}R_{\mathcal{C}}+\frac{2}{n}\log V(\Lambda')- \log \bigg(2\pi e \sigma^2 \frac{e^\zeta}{2\sigma_h^2} \bigg) .
\label{eqn:VNRfading}
\end{eqnarray}}
Define
\begin{equation}
\begin{cases} \epsilon_{1}=C_H(\Lambda,\sigma^2), \\
\epsilon_{2}=nE_h\Big[\mathfrak{h}\Big(\frac{\sigma^2}{h^2}\Big)\Big]-E_h\Big[\mathfrak{h}\Big(\Lambda' ,\frac{\sigma^2}{h^2}\Big)\Big], \\
\epsilon_{3}=C_H(\Lambda/\Lambda', \sigma^{2})-R_{\mathcal{C}}=\sum_{\ell=1}^{r}{C_H(\Lambda_{\ell-1}/\Lambda_{\ell}, \sigma^{2})-R_{\ell}},
\end{cases}
\label{eqn:epsilons}
\end{equation}
We note that, $\epsilon_{1}\geq0$ represents the capacity of the mod-$\Lambda$ fading channel, $\epsilon_{2}\geq0$ due to the data processing theorem, and $\epsilon_{3}\geq0$ is the total capacity loss of component codes.

Then we have
\begin{eqnarray}
\log \left(\frac{\gamma_{L}(\sigma)}{2\pi e} \cdot \frac{2\sigma_h^2}{e^\zeta}\right)
=\frac{2}{n}(\epsilon_{1}-\epsilon_{2}+\epsilon_{3}).
\end{eqnarray}
Since $\epsilon_2 \geq 0$, we obtain the upper bound
\begin{eqnarray}
\log \left(\frac{\gamma_{L}(\sigma)}{2\pi e} \cdot \frac{2\sigma_h^2}{e^\zeta}\right)
\leq \frac{2}{n}(\epsilon_{1}+\epsilon_{3}).
\label{eqn:minimumVNR}
\end{eqnarray}

By the design criteria (a)-(c), we have $\epsilon_{1} \to 0$ and $\epsilon_{3} \to 0$. Therefore, $\log \left(\frac{\gamma_{L}(\sigma)}{2\pi e} \cdot \frac{2\sigma_h^2}{e^\zeta}\right) \to 0$, which represents the Poltyrev capacity. The right hand side of \eqref{eqn:minimumVNR} gives an upper bound on the gap to the Poltyrev capacity of the ergodic fading channel.
\end{IEEEproof}

\begin{rem}
The slowly vanishing error probability $P_e(L,\sigma^2)=O\big(\frac{1}{N^{2\delta-1}}\big)$ is mainly caused by the uncoded error probability $E_h\Big[P_e\Big(\Lambda',\frac{\sigma^2}{h^2}\Big)\Big]$ associated with the bottom lattice $\Lambda'$. As we will see in the next section, a sub-exponentially vanishing error probability can be achieved when the power constraint is taken into consideration, because the probability of choosing a non-zero lattice point from $\Lambda'$ vanishes exponentially in the lattice Gaussian distribution.
\end{rem}

\subsection{Polar Lattices With Gaussian Shaping}
In this subsection, we discuss the lattice Gaussian shaping for the polar lattices constructed for fading channels. It is well known that shaping is a source coding problem merely related to the chosen input distribution. For the case in which only the receiver knows CSI, the optimal input distribution for fading channels is the continuous Gaussian distribution \cite{el2011network}, which is the same as that for AWGN channels. It has been shown in \cite{LingBel13} that lattice Gaussian distribution preserves many properties of the continuous Gaussian distribution, including the ability of achieving the AWGN capacity. Therefore, the lattice Gaussian shaping technique proposed for the AWGN-good polar lattices in \cite{polarlatticeJ} can be applied to the fading channel with minor modification.

It has been proved in \cite{LingBel13} that the lattice Gaussian distribution preserves the capacity of the
AWGN channel when the associated flatness factor is negligible.

\begin{ther}
[Mutual information of lattice Gaussian distribution \cite{LingBel13}]\label{thm:capacity}
Consider an AWGN channel $Y=X+Z$ where the input constellation $X$ has
a discrete Gaussian distribution $D_{\Lambda-{c},\sigma_s}$
for arbitrary ${c} \in \mathbb{R}^n$, and where the variance
of the noise $Z$ is $\sigma^2$. Let the average signal power be $P$, and let $\tilde{\sigma}\triangleq \frac{\sigma_s\sigma}{\sqrt{\sigma_s^2+\sigma^2}}$ be the minimum mean square error (MMSE) re-scaled noise deviation. Then, if
$\varepsilon = \epsilon_{\Lambda}\left(\tilde{\sigma}\right) < \frac{1}{2}$ and $\frac{\pi\varepsilon_t}{1-\epsilon_t}\leq \varepsilon$ where
\begin{equation}
\varepsilon_t \triangleq
\left\{
  \begin{array}{ll}
    \epsilon_{\Lambda}\left(\sigma_s/\sqrt{\frac{\pi}{\pi-t}}\right), & \hbox{$t \geq 1/e$} \\
    (t^{-4}+1)\epsilon_{\Lambda}\left(\sigma_s/\sqrt{\frac{\pi}{\pi-t}}\right), & \hbox{$0< t < 1/e$}
  \end{array}
\right.
\end{equation}
the discrete Gaussian constellation results in mutual information
\begin{equation}\label{eq:lattice-capacity}
I_D \geq \frac{1}{2}\log {\bigg(1+\frac{P}{\sigma^2}\bigg)} - \frac{5\varepsilon}{n}
\end{equation}
per channel use.
\end{ther}

Motivated by Theorem \ref{thm:capacity}, one may choose a low-dimensional $\Lambda$ such as $\mathbb{Z}$ and $\mathbb{Z}^2$ whose mutual information has a negligible gap to the AWGN channel capacity, and then construct polar lattices to achieve the capacity.

For the ergodic fading channel with power constraint $P$, letting the input $X$ be Gaussian, the ergodic channel capacity is given by \cite{el2011network}
\begin{equation}
\begin{aligned}
I(X;Y,H)&=E_h\bigg[\frac{1}{2} \log \bigg(1+\frac{Ph^2}{\sigma^2}\bigg)\bigg] \\
&=\frac{1}{2} \int_0^{\infty} \frac{h}{\sigma_h^2} e^{-\frac{h^2}{2\sigma_h^2}} \log \bigg(1+\frac{Ph^2}{\sigma^2}\bigg) dh\\
&=\frac{1}{2} \log e \int_{t=0}^{\infty} e^{-t} \ln \bigg(1+\frac{2\sigma_h^2P}{\sigma^2}t\bigg) dt \\
&=\frac{1}{2} \log e \cdot \exp \bigg(\frac{\sigma^2}{2\sigma_h^2P}\bigg) E_1\bigg(\frac{\sigma^2}{2\sigma_h^2P}\bigg),
\end{aligned}
\end{equation}
where $\frac{1}{2} \log \big(1+\frac{Ph^2}{\sigma^2}\big)$ is the capacity of an AWGN channel with noise variance $\frac{\sigma^2}{h^2}$ and the same power constraint. To achieve the ergodic capacity, our strategy is to pick a lattice Gaussian distribution which is able to achieve the AWGN capacity $\frac{1}{2} \log \big(1+\frac{Ph^2}{\sigma^2}\big)$ for almost all possible $h$. For an instant Gaussian noise variance $\frac{\sigma^2}{h^2}$, the MMSE re-scaled noise in Theorem \ref{thm:capacity} is now a function of $h$ and has standard deviation $\tilde{\sigma}(h)= \frac{\sigma_s\sigma}{\sqrt{h^2\sigma_s^2+\sigma^2}}$. For a component lattice $\Lambda$, by Remark \ref{rk:flatness}$, \epsilon_{\Lambda}\left(\tilde{\sigma}(h)\right)$ increases as $h$ grows. We can choose such that $\varepsilon=\epsilon_{\Lambda}\left(\tilde{\sigma}(h_l)\right) \to 0$ for a large $h_l$, then the resulted mutual information by $D_{\Lambda-{c},\sigma_s}$ is lower-bounded as
\begin{equation}
\begin{aligned}
E_h[I_D(h)]&=\int_{h=0}^{h_l} P_H(h) I_D(h) dh+ \int_{h_l}^{\infty} P_H(h) I_D(h) dh \\
& \geq \int_{h=0}^{h_l} P_H(h) \bigg(\frac{1}{2}\log {\bigg(1+\frac{P h^2}{\sigma^2}\bigg)} - \frac{5\varepsilon}{n}\bigg) dh \\
& \geq E_h\bigg[\frac{1}{2} \log \bigg(1+\frac{Ph^2}{\sigma^2}\bigg)\bigg]- \int_{h_l}^{\infty} \frac{1}{2}P_H(h) \frac{Ph^2}{\sigma^2} dh -\frac{5\varepsilon}{n},\\
&=E_h\bigg[\frac{1}{2} \log \bigg(1+\frac{Ph^2}{\sigma^2}\bigg)\bigg]-\frac{1}{2} \bigg(\frac{h_l^2}{\sigma_h^2}+2\bigg)\frac{P\sigma_h^2}{\sigma^2} e^{-\frac{h_l^2}{2\sigma_h^2}}-\frac{5\varepsilon}{n}.
\end{aligned}
\end{equation}

Let $c=0$ for simplicity. For sufficiently large $h_l$ and small $\varepsilon$, $D_{\Lambda,\sigma_s}$ is able to approach the ergodic capacity. Let the binary partition chain $\Lambda/\Lambda_{1}/\cdots/\Lambda_{r-1}/\Lambda'/\cdots$ be labelled by bits $X_{1},\cdots,X_{r},\cdots$. Then, $D_{\Lambda,\sigma_s}$ induces a distribution $P_{X_{1:r}}$ whose limit corresponds to $D_{\Lambda,\sigma_s}$ as ${r\rightarrow\infty}$.

By the chain rule of mutual information
\begin{eqnarray}\label{eqn:chainrule}
I(Y,H;X_{1:r})=\sum_{\ell=1}^{r} I(Y,H;X_\ell|X_{1:\ell-1}),
\end{eqnarray}
we obtain $r$ binary-input channels ${W}_{\ell}$ for $1\leq \ell \leq r$. Given $x_{1:\ell-1}$, denote by $\mathcal{A}_\ell(x_{1:\ell})$ the coset of $\Lambda_{\ell}$ indexed by $x_{1:\ell-1}$ and $x_{\ell}$. Similar to \cite[eq. (17)]{polarlatticeJ}, the channel transition PDF of the $\ell$-th channel ${W}_{\ell}$ is written as
{\allowdisplaybreaks\begin{eqnarray}\label{eqn:transition}
&&\hspace{-5em}P_{Y,H|X_\ell,X_{1:\ell-1}}(y,h|x_\ell,x_{1:\ell-1}) \notag \\
&=&\text{exp}\left(-\frac{\|\frac{y}{h}\|^2}{2(\sigma_s^2+\frac{\sigma^2}{h^2})}\right)\frac{P_H(h)}{f_{\sigma_s}(\mathcal{A}_\ell(x_{1:\ell}))}\frac{1}{2\pi\sigma\sigma_s}\sum_{a\in \mathcal{A}_\ell(x_{1:\ell})}\text{exp}\left(-\frac{\|\alpha(h) y-a\|^2}{2{\tilde{\sigma}^2(h)}}\right),
\end{eqnarray}}where $\alpha(h)=\frac{h\sigma_s^2}{h^2\sigma_s^2+\sigma^2}$ and $\tilde{\sigma}(h)=\frac{\sigma_s\sigma}{\sqrt{h^2\sigma_s^2+\sigma^2}}$ are the generalized MMSE coefficient and noise standard deviation. In general, ${W}_{\ell}$ is asymmetric, and we have to employ the polar coding technique for asymmetric channels \cite{aspolarcodes} to achieve the capacity $I(Y,H;X_\ell|X_{1:\ell-1})$ of each level.

As shown in \cite{polarlatticeJ}, the construction as well as the decoding of polar codes for a BMAC can be converted to that for a BMSC by channel symmetrization (see \cite[Lemma 7]{polarlatticeJ}). By replacing $Y$ with $(Y,H)$, \cite[Th. 5]{polarlatticeJ} and \cite[Th. 6]{polarlatticeJ} can be easily extended to our work. Therefore, the construction method of multilevel polar codes in \cite{polarlatticeJ} works for the fading case as well. Besides information bits and frozen bits at each level, we have shaping bits which are determined by the former two according to the lattice Gaussian distribution. Applying a similar argument as in \cite[Lemma 10]{polarlatticeJ}, the symmetrized channel of $W_\ell$ at each level is equivalent to a $\Lambda_{\ell-1} / \Lambda_{\ell}$ fading channel. Consequently, the resultant polar codes for the symmetrized channels are sequentially nested by the analysis in Sect. \ref{sec:notPower}, and hence we obtain a polar lattice $L$ which is Poltyrev capacity-achieving for the i.i.d. fading channel. Moreover, the multistage decoding is performed on the MMSE-scaled signal $\alpha(h) y$ (cf. \cite[Lemma 8]{polarlatticeJ}). Since the frozen sets of the polar codes are filled with random bits (but shared with the receiver), we actually obtain a coset $L+{c}'$ of the polar lattice, where the shift ${c}'$ accounts for the effects of all random frozen bits. Finally, since we start from $D_{\Lambda,\sigma_s}$, we would obtain $D_{\Lambda^N,\sigma_s}$ without coding; since $L+{c}' \subset \Lambda^N$ by construction, we obtain a discrete Gaussian distribution $D_{L+{c}',\sigma_s}$.

With regard to the number of partition levels, the same analysis given in Sect. \ref{sec:notPower} can be applied. By setting $h_l=O(N)$ and $\Lambda=\frac{1}{h_l}\Lambda_f$ for a fine lattice $\Lambda_f$, we have $\mathfrak{h}\big(\Lambda, \tilde{\sigma}^2(h_l)\big) \approx \log V(\Lambda)$ and hence $\epsilon_{\Lambda}(\tilde{\sigma}(h_l)) \to 0$ as $N \to \infty$ by the same argument of \eqref{eq:EHapprox}. Note that $\tilde{\sigma}(h_l) \to \frac{\sigma}{h_l}$ for large $h_l$. However, for the small channel gain $h_s$, we do not need $h_s=O\big(\frac{1}{N^\delta}\big)$ because of the lattice Gaussian shaping. To see this, let $h_s=1$ and the bottom lattice $\Lambda'=\Lambda_c$ for a coarse lattice $\Lambda_c$. By the definition \eqref{eq:latticeGaussian} of lattice Gaussian distribution, the probability of choosing a lattice point which is outside of $\mathcal{V}(\Lambda')$ is given by
\begin{eqnarray}
\begin{aligned}
\sum_{\lambda \in \Lambda' \setminus \{0\}} D_{\Lambda',\sigma_s}(\lambda)&=
\sum_{\lambda \in \Lambda_c \setminus \{0\}} D_{\Lambda_c,\sigma_s}(\lambda)\\
&=\sum_{\lambda \in \Lambda_c \setminus \{0\}}\frac{f_{\sigma_s}(\lambda)}{\sum_{\lambda' \in \Lambda_c}f_{\sigma_s}(\lambda')} \\
&\leq \frac{\sum_{\lambda \in \Lambda_c \setminus \{0\}}f_{\sigma_s}(\lambda)}{f_{\sigma_s}(\lambda'=0)} \\
&\leq (\sqrt{2 \pi}\sigma_s)^n P_e(\Lambda_c,\sigma_s^2)\\
&\leq n(\sqrt{2 \pi}\sigma_s)^n Q\big(\frac{d}{2\sigma_s}\big)\\
&\leq n (\sqrt{2 \pi}\sigma_s)^n \exp\big(-\frac{d^2}{8\sigma_s^2}\big).
\end{aligned}
\end{eqnarray} Recall that the minimum distance $d$ of $\Lambda_c$ scales as $d=O(\sqrt{N})$, and the second inequality satisfies for sufficiently large $d$ \footnote{Taking $n=1$ for an example, it is easy to check that $\sum_{\lambda \in \Lambda_c \setminus \{0\}}f_{\sigma_s}(\lambda) \leq 2 \int_{d-1}^{\infty} f_{\sigma_s}(x) dx \leq 2 \int_{\frac{d}{2}}^{\infty} f_{\sigma_s}(x) dx=P_e(\Lambda_c, \sigma_s^2)$ for a sufficiently large $d$. A similar argument holds for higher dimensions.}. Then $\sum_{\lambda \in \Lambda' \setminus \{0\}} D_{\Lambda',\sigma_s}(\lambda)$ vanishes exponentially for a fixed $n$ and a sufficiently large $N$, which means that only one lattice point in $\mathcal{V}(\Lambda')$ is chosen with probability close to 1, and the lattice point from $\Lambda'$ can be directly decoded according to the lattice Gaussian distribution. Therefore, the uncoded error probability $E_h\Big[P_e\Big(\Lambda',\frac{\sigma^2}{h^2}\Big)\Big]$ associated with the bottom lattice $\Lambda'$ vanishes exponentially, and it can be ignored since the error probability of polar codes for each partition channel vanishes sub-exponentially. By the same argument of \eqref{eq:levelr}, the number of levels is given by
\begin{eqnarray}
\begin{aligned}
r&=\log \big( V(\Lambda')/V(\Lambda)\big) \\
&= n \log (h_l) + \log \big( V(\Lambda_c)/V(\Lambda_f)\big) \\
&=nO(\log N),
\end{aligned}
\end{eqnarray}which is sufficient to achieve the ergodic capacity.

We summarize our main result in the following theorem:
\begin{ther}\label{thm:main}
For a sufficiently large channel gain $h_l=O(N)$, choose a good constellation with negligible flatness factor $\epsilon_{\Lambda}(\tilde{\sigma}(h_l))$ and negligible $\epsilon_t$ as in Theorem \ref{thm:capacity}, and construct a polar lattice with $r=nO(\log N)$ levels. Then, for i.i.d. fading channels, the message rate approaches the ergodic capacity $E_h\Big[\frac{1}{2} \log \Big(1+\frac{Ph^2}{\sigma^2}\Big)\Big]$, while the error probability under the multi-stage decoding is bounded by
\begin{eqnarray}
P_{e}\leq r N 2^{-N^{\beta'}}, \quad 0<\beta'<0.5,
\label{eqn:errorboundshape}
\end{eqnarray}
as $N\to \infty$.
\end{ther}

\begin{IEEEproof}
The proof of Theorem \ref{thm:main} can be adapted from the proofs of \cite[Th. 5]{polarlatticeJ} and \cite[Th. 6]{polarlatticeJ} by replacing $Y$ with $(Y,H)$.
\end{IEEEproof}

\begin{figure}[ht]
    \centering
    \includegraphics[width=15cm]{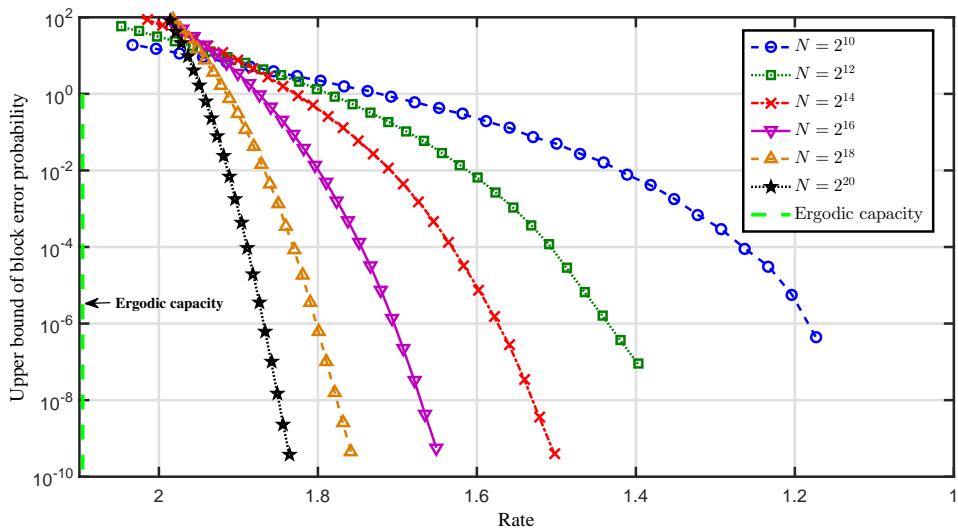}
    \caption{The upper-bounds of the block error probability of polar lattices under the SC decoding when $\sigma_s=3$, $\sigma=1$, $\sigma_h=1.2575$ and $N=2^{10}, 2^{12},..., 2^{20}$.}
    \label{fig:upperbound}
\end{figure}

Basing on the union bound, the upper-bounds of the block error probability of polar lattices under the SC decoding are plotted in Fig. \ref{fig:upperbound}, where $\sigma_s=3$, $\sigma=1$, $\sigma_h=1.2575$ and $N=2^{10}, 2^{12},..., 2^{20}$. Here we choose the binary-partition chain $\mathbb{Z}/2\mathbb{Z}/4\mathbb{Z}/8\mathbb{Z}/16\mathbb{Z}/32\mathbb{Z}$, and let $r=5$. In this case, the ergodic capacity is $2.0967$, and the channel capacities from level $1$ to level $5$ are given by $0.1213$, $0.5105$, $0.8437$, $0.5859$ and $0.0307$, respectively. Note that the gap between the achievable rate and the ergodic capacity is smaller than 0.2 for a block error probability $10^{-5}$ when $N=2^{20}$.

\section{Conclusion}
Explicit construction of polar codes and polar lattices for i.i.d. fading channels is proposed in this paper. By treating the channel gain as part of channel outputs, the work of polar codes and polar lattices for time-invariant channels is generalized to fading channels. We propose a simple construction of polar codes to achieve the ergodic capacity of binary-input i.i.d. fading channels when the CSI is not available to the transmitter. Furthermore, polar codes are extended to polar lattices to achieve the ergodic capacity of i.i.d. fading channels with certain power constraint.

\section*{Acknowledgments}
The authors would like to thank Dr. Xin Kang and Dr. Antonio Campello for helpful discussions and comments.

\bibliographystyle{IEEEtran}
\bibliography{Myreff}
\end{document}